\documentclass[pre,aps,superscriptaddress,twocolumn]{revtex4-1} 

\usepackage{amssymb}
\usepackage{amsmath}
\usepackage{bbold}
\usepackage{braket}
\usepackage{graphicx}
\usepackage{comment}
\usepackage{xcolor}
\graphicspath{{figures/}}

\newcommand{\beq}{\begin{equation}}
\newcommand{\eeq}{\end{equation}}

\newcommand{\p}{\partial}

\begin{document}

\title{The effects of curvature on the propagation of undulatory waves in lower dimensional elastic materials}

\author{Jonathan Kernes} 
\affiliation{Department of Physics and Astronomy, UCLA, Los Angeles California 90095-1596, USA}

\author{Alex J. Levine}
\affiliation{Department of Physics and Astronomy, UCLA, Los Angeles California 90095-1596, USA}
\affiliation{Department of Chemistry and Biochemistry, UCLA, Los Angeles California 90095-1596, USA}
\affiliation{Department of Computational Medicine, UCLA, Los Angeles California 90095-1596, USA}

\date{\today}

\begin{abstract}
The mechanics of lower dimensional elastic structures depends strongly on the geometry of their stress-free state. Elastic deformations separate into in-plane stretching and lower energy
out-of-plane bending deformations. For elastic structures with a curved stress-free state, these two elastic modes are coupled 
within linear elasticity. We investigate the effect of that curvature-induced coupling on wave propagation in lower dimensional elastic structures, focusing on the 
simplest example -- a curved elastic rod. We find that the dispersion relation of the waves becomes gapped in the presence of finite curvature; bending modes are 
absent below a frequency proportional to the curvature of the rod. By studying the scattering of undulatory waves off regions of uniform curvature, 
we find that undulatory waves with frequencies in the gap associated with the curved region tunnel through that curved region 
via conversion into compression waves.  These results should be directly
applicable to the spectrum and spatial distribution of phonon modes in a number of curved rod-like elastic solids, including carbon nanotubes and biopolymer filaments. 
\end{abstract}
\maketitle

\section{introduction}

Lower dimensional elastic structures are materials in which one (or more) of their characteristic length scales is microscopic, 
while the others are not.  Examples include biopolymer filaments~\cite{broedersz2014} (two such microscopic lengths and 
one macroscopic one), ribbons~\cite{kovsmrlj2016ribbon}, and membranes or shells (one microscopic length and 
two macroscopic ones)~\cite{nelson2004statistical}.  The physics of lower dimensional elastic structures is broadly applicable to 
problems ranging from nanometer lengths in carbon nanotubes~\cite{wang2005flexural,bower1999deformation} to $\sim 10^{6}$ 
meters when discussing continental plates~\cite{kearey2009global}.  In the purely biological context, lower dimensional elastic structures 
are central to several systems, including viral capsids~\cite{lidmar2003virus,michel2006nanoindentation,klug2006failure,singh2020finite},
and cell membranes~\cite{waugh1979thermoelasticity,park2010measurement,park2011measurement}, as well as filaments and and their bundles. 

Due to their having one (or more) microscopic dimensions, lower dimensional elastic structures have a large separation of energy scales 
associated with deformation along the ``thin'' directions as compared to the directions normal to them~\cite{Landau1986}.  
This is well known in the study of flat elastic shells, in which the 
out-of-plane motion of the sheet, that arises due to bending deformations, requires low energies when compared to in-plane
deformations.  For a shell of lateral extent $L$ and thickness $h \ll L$, this separation of energy scales can be parameterized by the 
F\"{o}ppl-von K\'{a}rm\'{a}n number $\nu K \sim \left(L/h \right)^{2} \gg 1$~\cite{niordson2012shell,sanders1963nonlinear}.  
For a flat shell and within linear elasticity theory,
these soft bending modes decouple from the stiff in-plane deformations.  When the 
elastic reference (stress-free) state of the shell is not flat, these modes are coupled by the local curvature.  The result is that shells 
with complex geometry have significantly different elastic behavior~\cite{lazarus2012geometry,bende2015geometrically,kovsmrlj2013mechanical}. 
For example, thin shells with local positive Gauss curvature in their stress-free state 
inhibit bending undulations~\cite{vaziri2008localized}.   Previous studies of the dynamics of undulatory waves on curved shells have 
shown that, in the geometric optics limit, these waves are reflected and refracted by changes in the local curvature.  They can even 
undergo total internal reflection when propagating from regions of negative to positive Gaussian curvature~\cite{evans2013reflection}. Such
effects have measurable implications for the spatial 
distribution of thermal undulations on red blood cells, which have regions of both positive and negative
Gauss curvature~\cite{evans2017geometric}.

The coupling of bending to stretching by curvature alters the normal-mode frequency spectrum by 
mixing in-plane and out-of-plane deformations.  One may ask whether one could, in effect, ``hear'' the curvature of a shell by examining its 
eigenfrequencies of vibration.  Famously, such a question was posed with regard to hearing the shape of drum~\cite{kac1966can}, which 
was in the negative~\cite{gordon1992}.  We suggest by an example discussed below, that one can, in fact, hear the shape of a bent rod; this has implications for understanding the phonon structure of some carbon nanotubes~\cite{barnard2019real}.

In this manuscript, we study the propagation of elastic waves on an undamped filament, where the elastic reference state couples 
bending and stretching deformations within the framework of linear elasticity.  Our goal is go beyond the geometric optics 
analysis of undulatory waves and produce the analog of the Fresnel equations, allowing one to understand the transmission and reflection 
of elastic energy intensity at various geometric interfaces.  The simplest model that retains the geometric coupling of in-plane deformation
and bending is the elastic rod.  While we believe that these results will inform work on membranes with more complex curvature, the
theory is directly applicable to a wide variety of filaments.  After introducing the elastic Hamiltonian in Sec.~\ref{sec: model}, we 
analyze in Sec.~\ref{sec: modes} 
the effect of uniform curvature on the eigenmodes of a rod, addressing the question of whether one can, in this instance, observe
the effect of curvature on the mode spectrum. In Sec.~\ref{sec: scattering}, we look at the 
scattering of elastic waves on an infinite rod by localized regions of curvature, where we 
find that undulations can tunnel through curved regions that do not support such undulations in the bulk.  Finally, we summarize our results and 
comment on their implications in Sec.~\ref{sec: conclusion}.

\section{model}
\label{sec: model}
We consider the elastic dynamics of a thin curved rod embedded in two dimensions.  We neglect twisting/torsional modes of deformation.  
We do not consider the elastic deformation of the material in the rod's cross section. Where applicable, we will state the
results for a rod of uniform cross section and composed of isotropic elastic continuum with uniform elastic constants.

\begin{figure}
\includegraphics[width=1\linewidth]{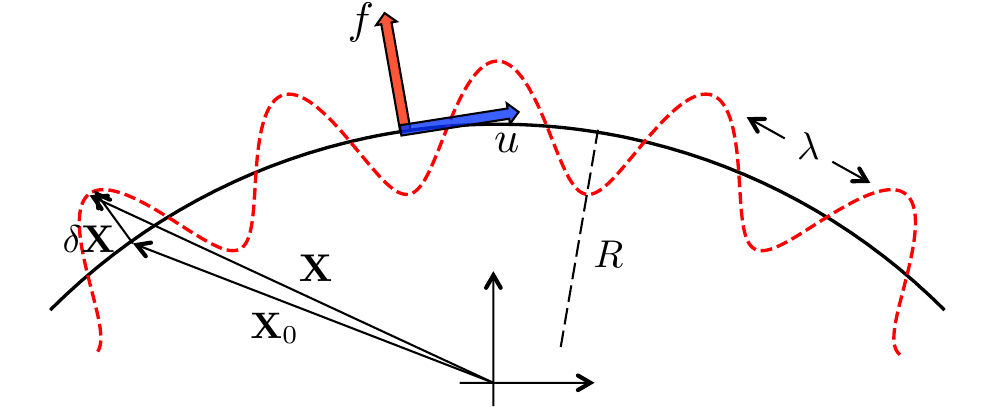}
\caption{(color online) Schematic representation of an undulatory wave on a curved rod. The (black) solid
 line is the space curve of the undeformed rod with radius of curvature $R$
supporting a sinusoidal wave (not to scale) shown as the (red) dashed line.  Deformations about the undeformed state are 
decomposed locally into a displacement $u$ (wide blue arrow) along the local tangent, and a displacement $f$ (wide red arrow) 
along the local normal. The weak curvature approximation assumes that the radius of curvature $R$ of the stress-free state (solid black line) is much larger 
than wavelength $\lambda$ of characteristic deformations (dashed red line).}
\label{fig: schematic}
\end{figure}

We develop the mechanics of curved rods by determining the action, from which the equations of motion are derived. We work in the 
weak curvature limit shown schematically in Fig.~\ref{fig: schematic}. The weak curvature limit is equivalent to the inequalities 
$ h \ll \lambda \ll R$, where $h$ represents the cross sectional radius, $\lambda$ the length of characteristic deformations, and $R$ the 
local radius of curvature. This is a one-dimensional version of the linearized shallow shell theory approximations~\cite{niordson2012shell,evans2013reflection}.

The stress-free configuration of the rod, shown in Fig.~\ref{fig: schematic}, is described by a two-dimensional space curve ${\bf X}_0(s)$, 
where $s$ denotes the arclength. The local tangent is given by ${\bf \hat t} = d {\bf X}_0/ds$. One may also compute 
the local normal (and binormal, which is trivial for the rod embedded in the plane) vectors 
via the well-known Frenet-Serret relations~\cite{spivak1970comprehensive}. 
We may write these relations as
\beq
\frac{ d{\bf \hat t}}{ds} = \kappa(s) {\bf \hat n}, \qquad \frac{d {\bf \hat n}}{ds} = - \kappa(s) {\bf \hat t},
\eeq
where $\kappa(s)$ is the arclength dependent curvature (equivalent to the inverse radius of of curvature $R(s)$), 
${\bf \hat n}$ the local normal vector, and bold-face symbols refers to two-dimensional vectors.

We neglect self-intersections of the rod so that its elastic energy density is determined solely by its local state of deformation. For 
small deformations $\delta X(s)$, the space curve describing the deformed state is then
\beq
{\bf X}(s) = {\bf X}_0(s) + \delta {\bf X}(s).
\eeq
Translational invariance demands that the elastic energy, $\mathcal{U}$, be function of $\frac{d {\bf X}_0}{ds},\frac{d {\bf X}}{ds}$ 
and their derivatives. Given the curved stress-free state of the rod, we also require that
 $\mathcal{U}$ vanishes when ${\bf X}={\bf X}_0$. We obtain the elastic energy
\beq
\begin{split}
\mathcal{U} &= \frac{1}{2} \int ds \bigg[a U^2 + b K^2 \bigg],
\end{split}
\eeq 
where $a$ and $b$ represent phenomenological parameters governing stretching 
and bending respectively. We have introduced the one-dimensional longitudinal strain tensor
\beq
U =\frac{1}{\sqrt{2}}\left[ \left |\frac{d {\bf X}}{ds} \right|^2 - \left |\frac{d {\bf X}_0}{ds} \right|^2 \right]^{1/2} = {\bf \hat t} \cdot \frac{d \delta {\bf X}}{ds}+ \mathcal{O}(\delta X^2),
\eeq
and the bending tensor
\beq
K = \frac{1}{\kappa\sqrt{2}}\left[ \left |\frac{d^2 {\bf X}}{ds^2} \right|^2 - \left |\frac{d^2 {\bf X}_0}{ds^2} \right|^2 \right]^{1/2}=  {\bf \hat n} \cdot \frac{d \delta {\bf X}}{ds}+ \mathcal{O}(\delta X^2).
\eeq
Deformations are parametrized in normal coordinates~\cite{nelson2004statistical}
\beq
\delta {\bf X} = u {\bf \hat t} + f {\bf \hat n},
\eeq
where $u$ represents the local in-plane stretching, and $f$ the local out-of-plane bending. 
Rewriting $U$ and $K$ in terms of normal coordinates, we find (using primes to denote arclength derivatives)
\beq
\label{eq: U}
U = u' - \kappa f,
\eeq
and
\beq
\label{eq: K}
K = f'' + 2 \kappa u' + \kappa' u - \kappa^2 f.
\eeq
In the weak curvature limit, the second and fourth terms are negligible. The third term is more subtle. It can 
certainly be discarded for rods with constant curvature, which we study here, 
but also may be discarded provided $R'/R$ is small. We thus find $K \approx f''$.

To determine the action, we introduce the kinetic energy, taking the mass density (mass per unit length) of the rod 
to be $\rho$. In the weak curvature limit, the kinetic energy density can be 
approximated by its flat rod result, as corrections are higher order in curvature. Using dots (primes) for time (spatial) derivatives, 
we obtain the action
\beq
S = \frac{1}{2} \int ds \left\{ \rho \dot{f}^2 + \rho \dot{u}^2 - a (u'-\kappa f)^2 - b f''^2  \right\}.
\eeq
For a uniform elastic rod with Young's modulus $Y$, cross sectional area $A$, and moment of inertia $I$, the two 
phenomenological elastic constants can be expressed in terms of these more microscopic ones as: $a= Y A$ and $b = YI/2$~\cite{Landau1986}.  
We may eliminate the dependence on $a,b$ by a suitable rescaling of length and time, introducing dimensionless independent variables:
$s \rightarrow s/\ell^{*}$ and $ t \rightarrow t/t^{*}$, where 
\begin{eqnarray}
\ell^* &=&  \sqrt{\frac{b}{a}} \\
 t^* &=& \frac{\sqrt{b \rho}}{a}.
\end{eqnarray}

Variations with respect to $u$ and $f$ yield the equations of motion
\begin{subequations}
\label{eq: eom}
\beq
\label{eq: f eom}
\partial_t^2 f +\partial_s^4 f +  M^2 f = M \partial_s u\eeq
\beq
\label{eq: u eom}
\partial_t^2 u - \partial_s^2 u = -\partial_s(M f),
\eeq
\end{subequations}
where we have defined the dimensionless curvature 
\beq
M(s) = \ell^*/R(s),
\eeq
in terms of the the arclength dependent stress-free radius of curvature $R(s)$. Eqs.~\ref{eq: f eom} and~\ref{eq: u eom} 
are one-dimensional versions of the linearized shallow shell equations governing thin shells~\cite{niordson2012shell}.

The boundary conditions are also obtained by variation of the action. 
In addition to continuity of $u, \, f$, and $f'$ across the boundary, we find three force balance equations. These equations require the continuity 
\begin{eqnarray}
\label{eq: bcs}
\Delta(u' - M f) &=& 0 \\
\label{eq: bcs2}
\Delta(f'') &=&0 \\
\label{eq: bcs3}
\Delta(f''') &=&0,
\end{eqnarray}
across an interface where the curvature of the rod changes, say at $s=0$.  In the above equations we use the notation
$\Delta (\phi) = \lim_{s\rightarrow 0^{+}}\phi - \lim_{s\rightarrow 0^{-}}\phi$ to represent the discontinuity of some variable 
$\phi$ across a boundary. 
At  a boundary where the curvature changes discontinuously, there is a subtlety in that $\kappa'$ is not well defined, suggesting that we are not
justified in discarding the term $\kappa' u$  in the bending tensor $K$ -- see  Eq.~\ref{eq: K}. However, in the presence of discontinuous 
curvature, our assumptions leading to the derivation of $K$ cease to hold as well. Physically, the boundary conditions 
Eqs.~\ref{eq: bcs}--\ref{eq: bcs3} represent longitudinal force balance, transverse force balance, and torque balance across the interface. Within the 
linearized shallow shell theory~\cite{niordson2012shell} approximation $K \approx f''$, these boundary conditions still provide the correct physical continuity of force 
and torque.  Eqs.~\ref{eq: eom} and boundary conditions Eqs.~\ref{eq: bcs}--\ref{eq: bcs3} 
represent the minimal coupling of an elastic rod to curvature. 

\section{Eigenmodes and frequencies}
\label{sec: modes}
We consider the case of constant curvature, which corresponds to the replacement $M(s) \to M$. Eqs.~\ref{eq: f eom},~\ref{eq: u eom} 
now constitute a set of linear partial differential equations. In the frequency domain, these equations can be made to 
appear like the time-independent Schr\"{o}dinger equation for a spinor-valued state ket $\ket{\psi}$:
\beq
\label{eq: schrodinger}
\hat H \ket{\psi}=\omega^2 \ket{\psi},
\eeq
which may be written in the $s$ or arclength basis
\beq
\label{eq: psi s}
\braket{s | \psi} = f(s) \ket{f} + u(s) \ket{u},
\eeq
in terms of basis spinors $\ket{f} = (1\, 0)^{T}$ and $\ket{u} =(0 \,1)^{T}$, and two ``wavefunctions'' $f(s)$ and $u(s)$, which correspond to 
the amplitude of bending and stretching deformation respectively. 
In terms of this spinor $fu$ basis, the 
Hamiltonian is given by
\beq
\label{eq: H k}
\hat H =
\left(
\begin{array}{cc}
\p_s^4 + M^2 & - M \p_s \\
M \p_s & -\p_s^2
\end{array}
\right).
\eeq
Note that the $f$ and $u$ problems decouple on a straight rod ($M=0$) as expected -- see below.  
We look for traveling wave solutions of the form $e^{i k s} \ket{\psi_k}$, where the spinor $\ket{\psi_k}$ is $s$-independent. 
The Hamiltonian acting on such a state becomes
\beq
\label{eq: H}
\hat H(k,M) =\left(\begin{array}{cc}k^4 + M^2 & -iMk \\ i M k & k^2\end{array}\right).
\eeq

\subsection{zero curvature}
\label{sec: zero curvature}

We briefly review the case of zero curvature ($M=0$). The Hamiltonian is diagonal, with eigenfrequency and eigenstate pairs
\beq
(\omega=k) \leftrightarrow \ket{u} , \;\;\;   (\omega=k^2) \leftrightarrow \ket{f}.
\eeq
$f$ and $u$ waves have quadratic and linear dispersion relations respectively. In their mode spectrum there 
are  three points of degeneracy: $k=0,\pm1$.

For a finite rod of length $\ell$, boundary conditions restrict the allowed values of wavenumber $k$, producing a discrete spectrum 
of eigenvalues or frequencies. We consider a clamped, pinned rod, requiring that $u$, $f$, and $f'$ vanish at the boundary. 
The eigenvalue equation (Eq.~\ref{eq: schrodinger}) has a solution of the form $\ket{\psi(s)} = e^{i ks} \ket{\psi_k}$, 
provided that $\det(\omega^2\mathbb{1} -\hat H(k))=0$. This is satisfied for any wavenumber $k$ that fulfills the condition
\beq
(\omega^2-k^2)(\omega^2-k^4)=0.
\eeq
There are six solutions.  These include two propagating $u$ waves of the form $e^{\pm i \omega s}\ket{u}$, two propagating $f$ waves
of the form $e^{\pm i\sqrt{\omega} s} \ket{f}$, and two exponential (evanescent) $f$ waves corresponding to 
imaginary solutions of wavenumber. These are given by $e^{\pm \sqrt{\omega} s}\ket{f}$.
\begin{figure}
\includegraphics[scale=0.63]{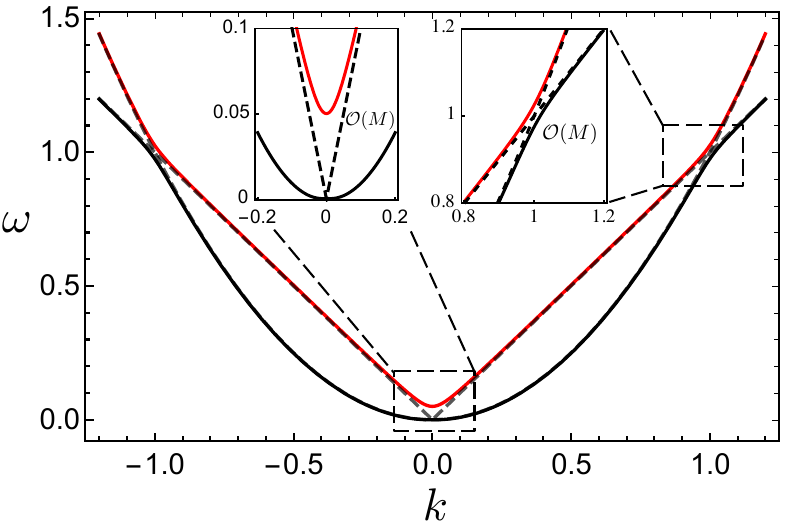}
\caption{(color online) Dispersion relation of a uniformly curved rod. $M=0.05$ and $\ell=1$. The degeneracy between the $M=0$ dispersion curves 
(dashed black lines) is lifted due to curvature. Level splitting between the upper branch (red) and lower branch (black) is $\mathcal{O}(M)$ 
near wavenumbers $k=\pm1$ and $k=0$.}
\label{fig: level splitting}
\end{figure}

We determine the allowed frequencies by first projecting $\ket{\psi}$ onto the wavenumber basis
\beq
\psi(k) = \sum_{\sigma=\pm} c_\sigma^u e^{\sigma i \omega s} \ket{u}+\left( c_\sigma^f e^{\sigma i \sqrt{\omega} s}+c_\sigma^{E,f} e^{\sigma \sqrt{\omega} s}\right) \ket{f},
\eeq
in terms of the undetermined coefficients $c_\pm^u,c_\pm^f,c_\pm^{E,f}$. The six boundary conditions (three at each end) 
produce a set of six equations for the six coefficients. A solution exists provided the determinant of the coefficient matrix vanishes, 
yielding the eigenfrequency condition
\beq
\left[\cos(\sqrt{\omega} \ell)\cosh(\sqrt{\omega} \ell) -1 \right]\sin(\omega \ell)=0.
\eeq
Frequencies $\omega_f$ that cause the bracketed expression to vanish correspond to purely $\ket{f}$ bending modes, 
whereas frequencies $\omega_u$ that cause the sine to vanish are purely $\ket{u}$ stretching modes. 
Since the function $\cosh(x)$ grows exponentially with its argument, to good approximation, we may use the 
approximate $f$-mode frequency condition $\cos(\sqrt{\omega_f} \ell)=0$ when $\sqrt{\omega} \ell > 1$.  This
leads to the (approximate) solutions for the bending-mode eigenfrequencies 
\beq
\label{eq: omega f}
\omega_f \approx \left(\frac{(n+1/2)\pi}{\ell}\right)^2,
\eeq
for positive integers $n$. The stretching eigenfrequencies, which correspond to vanishing of $\sin(\omega_u \ell)$, are easily
 found to be
\beq
\label{eq: omega u}
\omega_u= \frac{n \pi}{\ell}.
\eeq

\subsection{Uniform curvature}
In the presence of uniform curvature $M$, the eigenfrequencies of Eq.~\ref{eq: H} split into two branches:
\beq
\label{eq: omega pm}
\omega_\pm^2 = \frac{1}{2}\left[ (k^4+k^2+M^2) \pm \sqrt{(k^4+k^2+M^2)^2 - 4 k^6}\right],
\eeq
where the (+) subscript refers to the upper branch, and the (-) subscript to the lower. 
In the limit $M\to 0$ and $|k|>1$, these reduce to $\omega_+ = k^2$ and $\omega_- = k$, indicating that the upper branch corresponds to a 
bending mode, and the lower branch to a stretching mode. For $|k|<1$, the identification is reversed, with $\omega_+ =k$ and $\omega_- =k^2$. These identifications are further supported by looking at the 
(unnormalized) eigenmodes, which may be written as
\begin{eqnarray}
\label{eq: + ket}
\ket{+} &=& \ket{f} + \frac{- i k M}{k^2 - \omega_+^2(k,M)} \ket{u} \\
\label{eq: - ket}
\ket{-} &=& \ket{u} + \frac{i k M}{k^4 + M^2 - \omega_-^2(k,M) } \ket{f}.
\end{eqnarray}
For $|k|>1$, the $M \to 0$ limit recovers the zero-curvature results $\ket{+} = \ket{f}$ and $\ket{-} = \ket{u}$. Again, the identifications are reversed for $|k|<1$.
\begin{figure}
\includegraphics[scale=0.58]{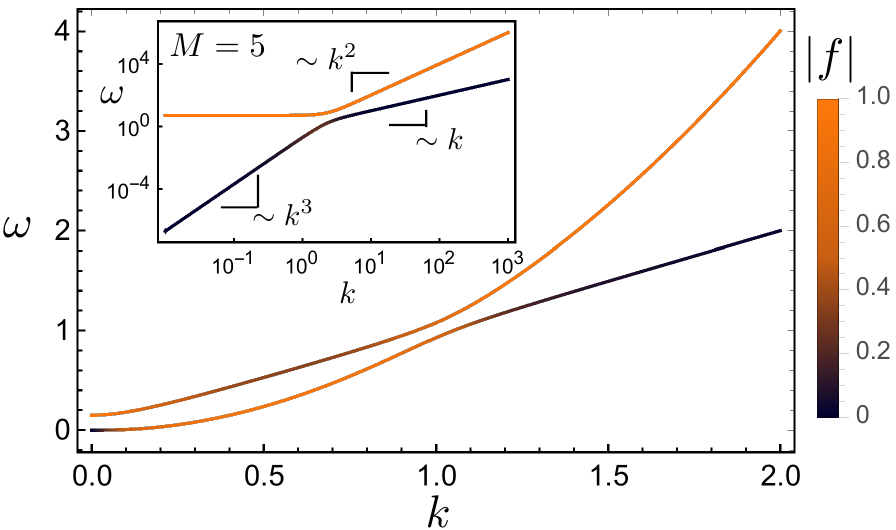}
\caption{(color online) The dispersion relation of a rod with uniform curvature, color-coded according to the normalized amplitude of its 
corresponding $\ket{f}$ eigenstate. $M=0.15$ and $\ell=1$. Mode mixing is strongest near the degeneracy points of the $M=0$ case. 
In the inset, we show the behavior for larger curvature $M>1$ and $\ell=1$. At small $k$, the upper branch is essentially flat, while the lower branch 
develops an $\omega_- \approx |k|^3/M$ power law, in contrast to its quadratic dispersion relation 
at small curvature.}
\label{fig: free dispersion}
\end{figure}

In Fig.~\ref{fig: level splitting}, we plot the dispersion relations of elastic waves on the rod at fixed $M \neq 0$. Curvature lifts the 
degeneracies at wavenumbers $k=0,\pm1$. The magnitude of the level splitting is $\mathcal{O}(M)$. The upper branch is gapped; it does not tend to zero with wavenumber, but instead to $\omega(k=0)=M$. In this sense, the upper branch acts as if it has acquired a {\it mass} due to curvature, hence our use of the notation $M$. If the system possess frequencies $\omega <M$, they must have complex wavenumber and are necessarily bound. At finite $M$, in the limit $k\to 0$, the eigenmode of the upper branch becomes a pure $\ket{ f}$ mode. Bending modes are gapped in the presence of curvature, which can be viewed as the one-dimensional analog of the suppression of undulations on thin shells at areas of positive Gauss curvature~\cite{vaziri2008localized,evans2013reflection}.

In the absence of curvature, the linear and quadratic dispersion corresponded to directly to $\ket u$ and $\ket f$ normal modes. In the presence of curvature, these normal modes are mixed. In Fig.~\ref{fig: free dispersion}, we show the same free dispersion relation color coded by normal mode amplitude. The amplitudes obey the normalization constraint $|u|^2 +|f|^2=1$, which implies that $|f|=1$ when $|u|=0$, and vice-versa. The effects of mode-mixing are most prevalent at wavenumbers near level splitting. At these points, the normal mode amplitudes of the two branches switch character between $u$ and $f$ dominated. This ensures that only bending (stretching) dominated normal modes exhibit quadratic (linear) dispersion at large wavenumber. For wavenumber $k <1$, the lower branch $\omega_- \sim |k|^3/M$, in contrast to the zero curvature quadratic dispersion.

At fixed $k$, the frequencies on the upper (lower) branch of Eq.~\ref{eq: omega pm} increase (decrease) with increasing $M$. 
At large $k$, the frequencies on the lower branch decrease $\sim M^{-1}$, while those on the upper branch are hyperbolic and approach
 the asymptote $\omega =M$. As a result, frequencies $\omega_+$ may never fall below $M$. This is due to the $k=0$ band gap shown 
 in  Fig.~\ref{fig: level splitting} (see the inset of that figure).

The large curvature limit of Eqs.~\ref{eq: + ket},~\ref{eq: - ket} shows that the eigenstates that mix bending and stretching modes 
once again decouple so that $\ket{+} \to \ket{f}$ and $\ket{-} \to\ket{u}$. Interestingly, this is the same result as for $M \to 0$. 
Since at large $M$ the $\ket{+}$ states become pure bending modes, we deduce that bending dominated modes may not have 
frequencies $\omega <M$. Moreover, by increasing the curvature, one can identify which eigenfrequencies are related to 
primarily bending (stretching) dynamics, by seeing if they increase (decrease) with $M$. At larger curvature, due to the frequency gap, these frequencies are separated by the line $\omega=M$.

We now turn from the case of an infinite rod to a finite one.  For a finite rod, we must impose boundary conditions at the ends, which 
generally lead to a quantized set of eigenfrequencies $\omega_n$.  To study how the frequency spectrum changes with respect to 
curvature, we fix the pinned and clamped boundary conditions at the ends and vary only the curvature $M$. 
Thus, we demand that $u,\, f,$ and $f'$ vanish at the endpoints $s=0,\ell$.

Following the steps of Sec.~\ref{sec: zero curvature} to determine the eigenfrequencies involves solving a 
cubic characteristic equation for $k^2$ as a function of $\omega$, followed by finding the roots of an analytically complicated 
transcendental equation. Instead, we compute the eigenfrequencies and eigenfunctions directly in position space numerically, 
using collocation methods on a Chebyshev grid~\cite{trefethen2000spectral}. The eigenmode amplitudes are
determined via numerical integration $\frac{\int_0^\ell |f|^2 ds}{\int_0^\ell (|f|^2 + |u|^2) ds}$, performed via quadrature.  

\begin{figure}
\includegraphics[width=0.9\linewidth]{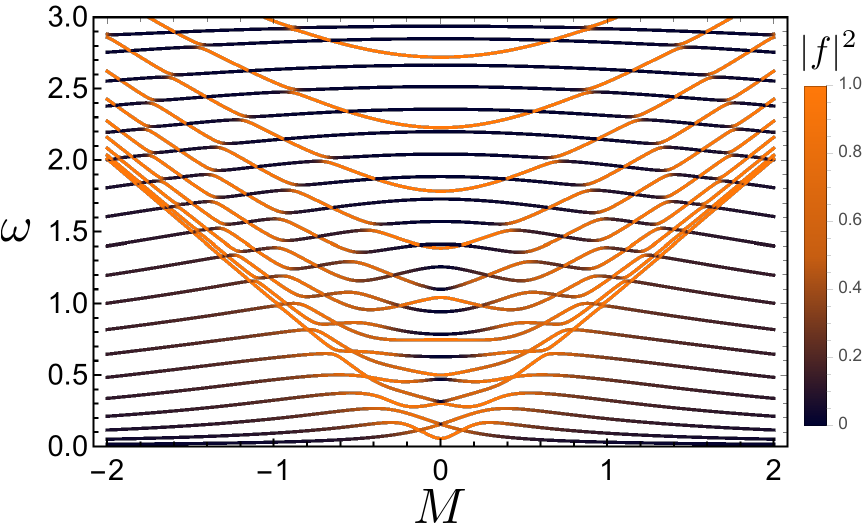}
\includegraphics[width=0.9\linewidth]{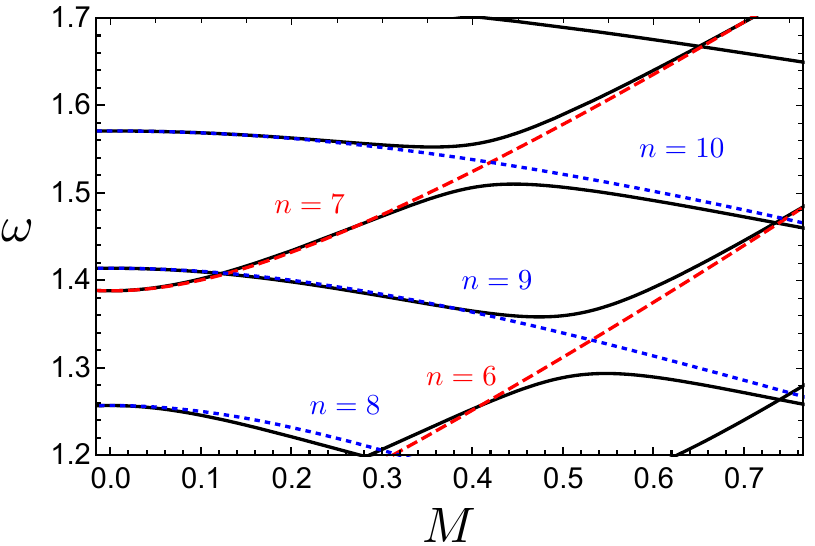}
\caption{(color online) The frequency spectrum of a clamped, pinned rod of length $\ell=20$ as a function of curvature $M$. 
Top: the $M$-dependence of the frequency spectrum, color coded by the relative amplitude of its $f$ to $u$ mode, where lighter colors 
represent more bending $f$-amplitude. We find three distinct regimes: 
high $\omega$ where the curves look like their infinite-rod counterparts, intermediate $\omega$, where they spectrum is approximated 
by free dispersion curves with level splitting, and low $\omega$ where curvature strongly distorts the spectrum. 
Bottom: a close up view of the frequency spectrum (black solid lines) 
overlaid with the infinite rod dispersion curves for several modes labeled by $n$ in the figure (dashed lines). Level splitting occurs between even
and odd numbered modes, as explained in the text.}
\label{fig: spectrum}
\end{figure}

In the upper panel of Fig.~\ref{fig: spectrum}, we plot the eigenfrequencies as a function of curvature for a rod of length $\ell=20$, 
color coded so that an increasing ratio of bending to stretching amplitude runs from dark to light. 
Broadly, frequencies that increase with respect to curvature are 
associated with bending $f$-modes, and such modes are still restricted to frequencies $\omega >M$. The lower frequency modes show more 
mixing of bending and stretching. 

Due to changing the rod's curvature, spectral lines (frequencies) corresponding to different modes cross. There are three regimes, 
dictated by the strength of interaction between different harmonics. At high frequency (and accordingly high $|k|$), curvature-induced 
coupling between bending and stretching is negligible. The spectral curves can be well approximated 
 by using the zero-curvature $k$ values, Eq.~\ref{eq: omega f} and Eq.~\ref{eq: omega u}, in the equations 
 for the $\omega_+$ and $\omega_-$ branches. As for the infinite rod, bending (stretching) modes increase (decrease) with increasing 
 curvature. At low frequencies, curvature significantly affects the rod, and the free dispersion relation gives a poor fit.

At intermediate frequencies (approximately $1<k<1.75$ in the upper panel of Fig.~\ref{fig: spectrum}), frequencies exhibit 
oscillatory behavior, due to level splitting between other harmonics. To understand this effect, we 
expand the state $\ket{\psi}$ of Eq.~\ref{eq: schrodinger} in the basis of zero curvature eigenmodes 
\beq
\label{eq: cn}
\ket{\psi} = \sum_n c_n(M) \ket{\psi_n^{(0)}},
\eeq
for some $M$-dependent coefficients. This leads to an equation for the coefficients $c_n$:
\beq
\left[\omega^2(M) - \omega_n^2\right] c_n = \sum_m \braket{ \psi_n^{(0)} | \hat V | \psi_m^{(0)} } c_m,
\eeq
where we have introduced the $M=0$ eigenfrequencies $\omega_{n}$ corresponding to the eigenmodes $\ket{\psi_{n}}$,
and we have defined the perturbation operator
\beq
\hat V = \left( \begin{array}{cc} M^2 & - M \partial_s \\ M \partial_s & 0\end{array} \right).
\eeq
The perturbed eigenfrequencies $\omega(M)$ retain implicit dependence on the curvature $M$.  For the infinite rod, 
the direct solution of Eq.~\ref{eq: cn} leads to the frequencies and states $\omega_\pm$, $\ket{\pm}$. We do not try to recover this result, 
but instead look at the possible straight-rod states coupled by the perturbation operator.
Evaluating the off diagonal matrix elements of $V_{mn}$ we find
\beq
\braket{m| \hat{V} | n}_{m \neq n} = -2 M \int_0^\ell f_m \partial_s u_n ds,
\eeq
where $f_m$ and $u_m$ represent the zero curvature eigenfunctions corresponding to the $m^{\rm th}$ and $n^{\rm th}$ 
eigenfrequencies -- see Eqs.~\ref{eq: omega f},~\ref{eq: omega u}. The zero-curvature Hamiltonian $\hat H_0$, is invariant under a
parity transformation $\hat H_0(s) = \hat H_0(-s)$. As a result, the eigenfunctions $f_n$ and $u_n$ are either even or odd. 
Since the operator $\partial_s$ is odd under parity, the operator $\hat V$ connects states of opposite parity. 
The coupling $V_{m\neq n}$ is non-vanishing only when $m$ is even and $n$ is odd, or vice-versa.
In the lower panel of Fig.~\ref{fig: spectrum}, we show a close-up view of the frequency spectrum overlaid with the free dispersion curves labeled by their harmonic. Level splitting occurs precisely between odd and even harmonics, which leads to the oscillatory-like behavior.

\section{scattering}
\label{sec: scattering}

We study the transmission and reflection of undulatory and compression waves through regions of nonzero curvature.  We 
imagine the scattering problem as follows.  
Two semi-infinite straight rod segments are appended to the left and right sides of a region of constant curvature M ({\em i.e.}~the arc of a circle), 
such that both the rod and its tangents are everywhere continuous.  We choose a coordinate
system so that the center $s=0$ is the symmetry point of the figure and note that the circular arc has length $\ell$.  The curvature jumps 
discontinuously from $0 \to M$ on the left, and $M \to 0$ on the right. See Fig.~\ref{fig: scatterpic}. 

In the straight domains $|s|>\ell/2$, waves are defined by the eigenmodes and eigenfrequencies of Sec.~\ref{sec: zero curvature}. 
Radiative incoming and outgoing states are thus determined solely by the basis of plane wave solutions, {\em i.e.}, values 
$k$ that satisfy the infinite rod dispersion relation. After demanding the solution be finite at $\pm \infty$, each 
semi-infinite rod has five such solutions: an incoming/outgoing $f$-wave, an incoming/outgoing $u$-wave, and one 
evanescent $f$-wave.

In the curved domain $s \in [-\ell/2,\ell/2]$, $k$ can take complex values. This differs from the well-known transmission through a 
barrier in quantum mechanics, where the allowed $k$ values are either purely real or imaginary~\cite{sakurai1995modern}. In general, 
states with real $k$ correspond to propagating solutions, and facilitate transmission. We refer to the number of propagating 
solutions in the curved region as the number of ${\it channels}$, whereby a wave may be transmitted through the curved domain. Before computing 
transmission/reflection coefficients for an incoming plane wave, we study how the number of available channels is set by the combination of 
both the curvature of the rod and the frequency of an incoming plane wave. 

The characteristic equation is found by demanding that the eigenvalue problem defined by Eq.~\ref{eq: schrodinger} with Hamiltonian given by Eq.~\ref{eq: H} has a solution. This is ensured provided 
\begin{equation}
\det \left[\omega^2 \mathbb{1} - \hat H(k,M)\right]=0.
\end{equation}
We find the characteristic equation for $\kappa =k^2$ is cubic:
\beq
\label{eq: characteristic-polynomial}
\kappa^3-\omega^2\kappa^2-\omega^2\kappa-\omega^2 \left(M^2-\omega^2 \right)=0.
\eeq
Real solutions $\kappa <0$ and $\kappa>0$ correspond to evanescent and propagating waves respectively. Complex $\kappa$ corresponds to damped propagating waves.

The number of channels is twice the number of real, positive roots $\kappa$. These roots are a function of frequency and curvature. 
At zero curvature ($M=0$), there are three roots at $\kappa = \pm \omega, \kappa = \omega^2$, leading to four channels 
(two pure $f$-waves, and two pure $u$-waves). At nonzero $M$, Descartes' rule of signs states that the number of positive (negative) 
roots is equal to or less than (by an even number) the number of sign changes of the 
coefficients when ordered in decreasing powers of $\kappa$ ($-\kappa$).

For $M>\omega$, the polynomial coefficients undergo one sign change. There is only one positive root. 
When $0\leq M <\omega$, Descartes' rule determines that there are either two or zero positive roots. 
In the limit $M=0$, we already know that the characteristic equation contains two positive roots, and has a positive $y$-intercept. Increasing 
$M$ will only serve to shift the characteristic polynomial downward, while keeping the $y$-intercept positive for $0\leq M<\omega$.  This shift 
cannot remove the two positive roots.  
We conclude that for $0 \leq M < \omega$ the characteristic polynomial has two positive roots.

In summary, there are four available channels when $0 \leq M < \omega$, but for $M > \omega$ there are only two available channels. The 
reduction in the number of channels with decreasing frequency can be traced back to the vanishing of $f$ dominated eigenmodes for 
frequencies $\omega <M$. However, the two available channels are not pure $u$-modes, but instead some combination of $f$ and $u$ 
displacements. This mixing of the modes allows pure $f$ or $u$ modes to interconvert in the presence of curvature, which has 
implications for phonon transmission through curved regions.

\subsection{Transmission/reflection through constant curvature}
\begin{figure}
\includegraphics[width=0.9\linewidth]{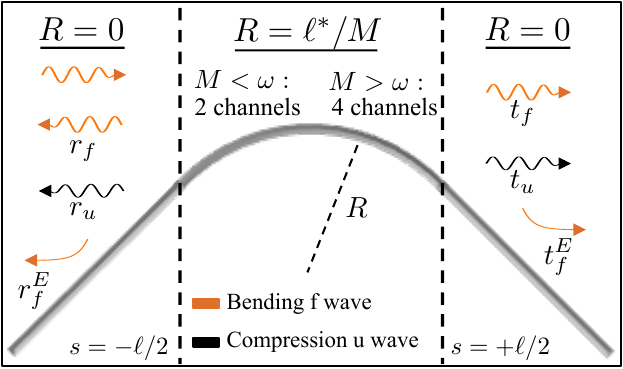}
\caption{(color online) Schematic representation of an elastic rod (solid gray line) formed by adjoining two semi-infinite straight 
rods at the (black) dashed lines to the left ($s=-\ell/2$) and right ($s=\ell/2$) of the curved rod segment (arc of a circle with radius $R$), such 
that the rod and its tangent are everywhere continuous. We consider the scattering of an incoming bending $f$ wave from the left, through the region of 
constant curvature $M$. Curvy arrows correspond to propagating asymptotic states, and decaying arrows to evanescent  states. The darker (lighter) colors refer to 
$u$ ($f$) modes. There are six unknown transmission/reflection amplitudes. In the curved region, there are either two or four propagating channels, 
determined by the value of $M$.}
\label{fig: scatterpic}
\end{figure}

We consider the case of an incoming, purely $f$ mode wave $e^{i\sqrt{\omega}s}\ket{f}$, or a purely $u$ mode wave $e^{i \omega s}\ket{u}$. In 
both cases, we take the incident wave to have unit amplitude far to the left of the circular arc. The wave, scattered by the curved region, 
produces two transmitted $f$ and $u$ waves with transmission amplitudes $t_f$ and $t_u$, two reflected waves with amplitudes $r_f$ and $r_u$, 
and two evanescent waves with amplitudes $r_f^E$ and $t_f^E$, which decay exponentially away 
from $s=\pm \ell/2$. The situation is summarized in Fig.~\ref{fig: scatterpic}.

The transmission (reflection) coefficient, denoted by a capital letter $T$ ($R$), is defined as the ratio of the 
outgoing flux of amplitude to the incoming flux. The flux is given by the product of the amplitude squared times the group velocity. 
For an incoming $f$ wave of unit amplitude, the $f$-mode transmission/reflection coefficients are 
\beq
T_f = |t_f|^2, \;\;\; R_f = |r_f|^2.
\eeq
However, since bending and compression waves obey different dispersion relations, we must account for their difference in group 
velocity. Compression $u$-waves have unit velocity, while bending waves have a group velocity of $d\omega/dk =2 \sqrt{\omega}$. 
For an incoming $f$ wave, the transmitted/reflected $u$ waves are given by 
\beq
T_u = \frac{|t_u|^2}{2\sqrt{\omega}}, \; \; \; R_u = \frac{|r_u|^2}{2\sqrt{\omega}}.
\eeq
To solve for the transmission/reflection coefficients, we must explicitly solve Eq.~\ref{eq: schrodinger} at nonzero $M$, 
and then employ the boundary conditions -- Eqs.~\ref{eq: bcs}-\ref{eq: bcs3} -- to stitch together solutions at the 
boundaries $s=\pm \ell/2$. 

Since we are looking for plane wave solutions of the form $\ket{\psi(s)} = e^{i k s} \ket{\psi_k}$, 
we shall reformulate Eq.~\ref{eq: schrodinger} as an eigenvalue problem of the operator $\partial_s$ at fixed $\omega$. 
This is accomplished by reducing all higher order derivatives $\p_s$ through the introduction of new 
fields $f_a \equiv \p_s^a f$ and $u_a = \p_s^a u$, for integers $a\geq 0$. The resulting system of 
equations may be written as a vector differential equation
\beq
\label{eq: A chi}
\partial_x \ket{\chi} =\hat A \ket{\chi}
\eeq
for six-dimensional vector
\beq
\label{eq: chi}
\ket{\chi} =\left(\begin{array}{c} u_0 \\ u_1 \\ f_0 \\ f_1 \\ f_2 \\ f_3\end{array}\right)
\eeq
and matrix
\beq
\label{eq: A}
\hat A =\left(
\begin{array}{cccccc}
 0 & 1 & 0 & 0 & 0 & 0 \\
 -\omega ^2 & 0 & 0 & M & 0 & 0 \\
 0 & 0 & 0 & 1 & 0 & 0 \\
 0 & 0 & 0 & 0 & 1 & 0 \\
 0 & 0 & 0 & 0 & 0 & 1 \\
 0 & M &\omega^2-M^2 & 0 & 0 & 0 \\
\end{array}
\right).
\eeq
The boundary conditions are now algebraic relations amongst all these fields.  For a trial solution of the form $e^{iks} \ket{\chi_k}$, there are six possible solutions
in the curved region, one for each value $k$ that is a root of the characteristic polynomial -- see Eq.~\ref{eq: characteristic-polynomial}.  The full solution is given 
by a linear superposition these trial solutions with six undetermined coefficients.  We also have six more undetermined coefficients associated with the incident, reflected, and 
transmitted waves, giving a total of twelve undetermined coefficients.  These are fixed by imposing the continuity of $u$, $f$, and $f'$ at $s=\pm \ell/2$ (six conditions), 
as well as the three force balance conditions --Eq.~\ref{eq: bcs} --at $s=\pm \ell/2$ (six conditions).  Thus, we have a system of twelve linear
equations that can be solved for the scattering amplitudes.

\begin{figure}
\includegraphics[width=0.9\linewidth]{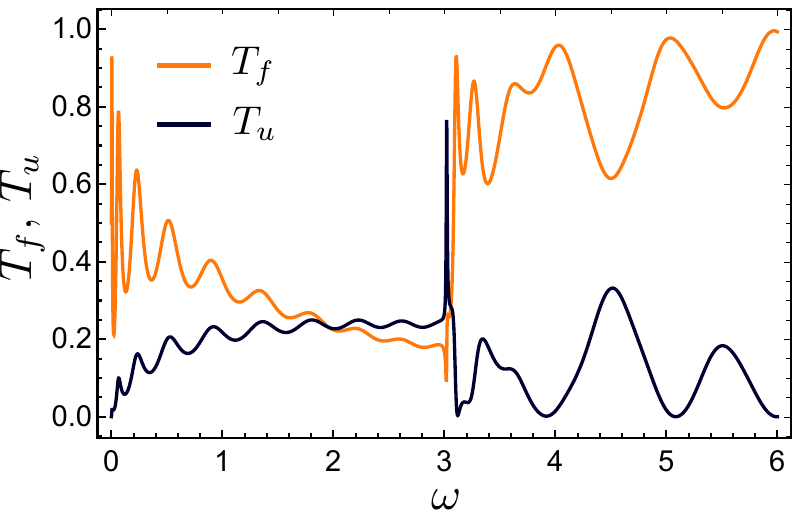}
\includegraphics[width=0.9\linewidth]{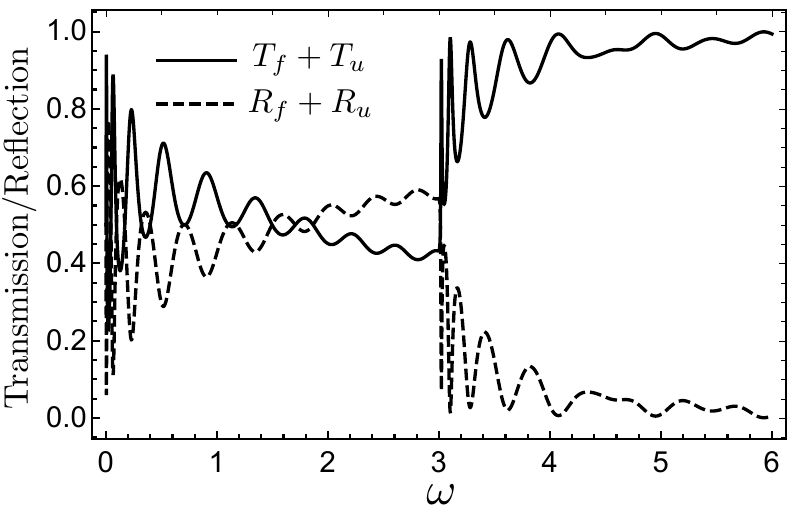}
\caption{(color online) Transmission and reflection coefficients for a bending wave of unit amplitude incident on a region of length $\ell=10$, with 
uniform curvature $M=3$. Top: the transmission coefficients for bending (orange) and compression (black) waves. Bottom: the total transmission and reflection coefficients. 
Due to conservation of energy, the coefficients obey $T_f+T_u+R_f+R_u=1$. Curvature mixes eigenmodes, converting the incident pure bending 
wave into a linear combination of bending and compression waves.}
\label{fig: transmission}
\end{figure}
We solve these equations numerically. In Fig.~\ref{fig: transmission}, we plot the transmission and reflection coefficients for an $f$ wave 
of unit amplitude incident on an interval of length $\ell=10$, with uniform curvature $M=3$. 
In the upper panel of Fig.~\ref{fig: transmission}, we show both the bending and compression transmission coefficients separately. 
Due to conservation of energy, we can define a total transmission coefficient $T_\text{tot} = T_f + T_u$, 
and reflection coefficient $R_\text{tot}=R_f+R_u$, such that their sum $T_\text{tot}+R_\text{tot}=1$ is unity. Though we consider 
only an incoming bending wave, we find that curvature allows the rod to convert bending into stretching deformations, leading to the production of compression waves.

At low frequencies, $\omega<M$, the circular arc of the rod cannot support bending-dominant modes. The nonzero transmission coefficient for incident bending waves indicates that the $f$-waves can, in effect, tunnel through the curved region via conversion to 
compression $u$-waves, which then convert back into outgoing bending $f$-waves in the righthand straight segment of the
rod. In the curved domain, the incoming bending mode propagates through one of the two available channels. As frequency increases through
$\omega=M$, the number of available channels in the curved domain jumps from two to four. This leads to a dramatic increase in 
the transmission coefficient.

For higher frequencies,  $\omega >M$, the circular arc can support bending-dominant modes. As a result, the transmission coefficient for 
$f$-waves in the upper panel of Fig.~\ref{fig: transmission} is much larger than that for $u$-waves, and tends to one as $\omega \to \infty$. 
The two principal effects of curvature -- conversion from bending to compression and suppression of bending modes -- 
diminish at high frequency.

In addition to these jumps, the transmission coefficients are oscillatory. It is well-known that peaks in the scattering amplitude 
correspond to bound states under a change of sign of the eigenvalue $\omega^2 \to -\omega^2$~\cite{gottfried2013quantum}.  Since $f$ has these peaks, 
they must correspond to eigenmodes in the curved region, which we know to be $u$-dominant. Therefore, the incident bending wave uses these $u$-dominant 
modes to ``tunnel'' through the curved region. 

When $\omega<M$, the upper panel of Fig.~\ref{fig: transmission} shows $T_f$ and $T_u$ oscillating in phase. This supports 
idea that bending modes propagate via compression-dominated eigenmodes in the curved domain. Alternatively,  for $\omega >M$, $T_f$ and $T_u$ oscillate out of phase. 
Peaks in $T_f$ occur at frequencies corresponding to bending-dominated bound states. The fact that $T_f$ and $T_u$ are now out of phase shows that 
bending $f$ waves are not traversing the curved region 
by conversion into compression $u$ waves.  

Finally, we observe that $T_f$ is a decreasing function of frequency in the domain $0\leq \omega <M$, 
while $T_u$ is an increasing function of frequency on that same domain. Transmission of bending waves is a minimum for frequencies 
just below $\omega =M$. This suggests that bending waves are most effective at tunneling through curvature for both small 
and large frequencies.

\begin{figure}
\includegraphics[width=0.9\linewidth]{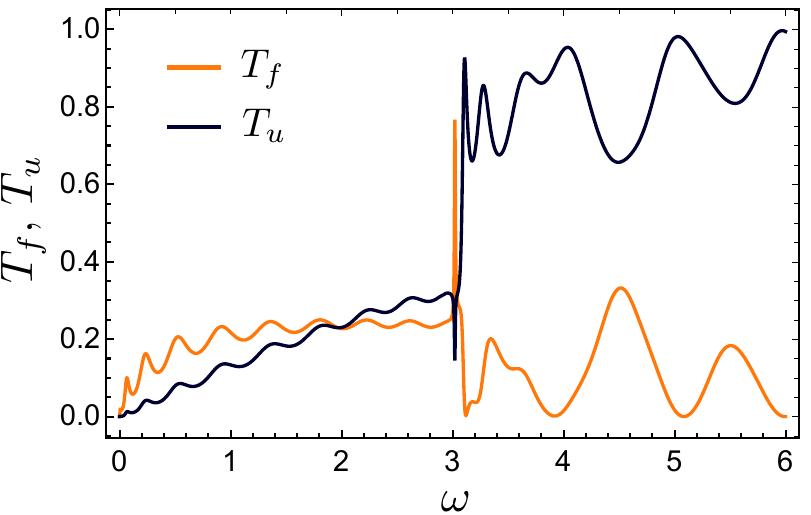}
\includegraphics[width=0.9\linewidth]{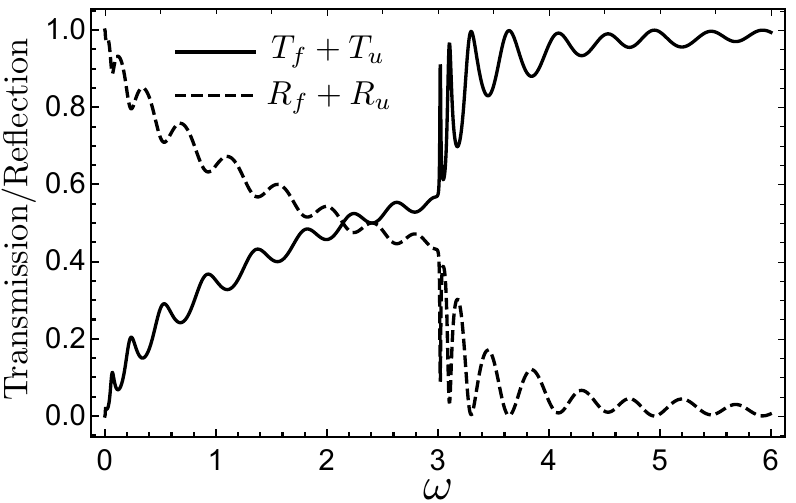}
\caption{(color online) Transmission and reflection coefficients for a compression wave of unit amplitude incident on a region of length $\ell=10$, with 
uniform curvature $M=3$. Top: the transmission coefficients for bending (orange) and compression (black) waves. Bottom: the total 
transmission and reflection coefficients. In contrast to an incident bending wave, see Fig.~\ref{fig: transmission}, the transmission 
coefficients vanish as $\omega \to 0$.}
\label{fig: transmissionU}
\end{figure}

In Fig.~\ref{fig: transmissionU} we show the transmission and reflection coefficients for the case of an incoming $u$-wave. We find 
similar results. The main difference lies at frequencies $\omega <M$. Bending waves arise only if they are produced via mode 
coupling in the curved domain. We find that $T_f$ follows $T_u$, decreasing as frequency goes to zero, in contrast to its behavior 
for a purely bending incoming wave.

\section{conclusion}
\label{sec: conclusion}

We investigate the interplay of bending and stretching in a curved,  one-dimensional elastic rod. This is the simplest model 
that retains both bending and stretching deformations, and allows their coupling via the geometry of the unstressed state. 
In the limit of small deformations, we find a set of two coupled equations for out-of-plane deformations $f$ and in-plane deformations $u$, 
corresponding to bending and stretching respectively.  These equations are the one-dimensional analog of the linearized shallow shell equations for a thin 
elastic shell.  In fact, those equations reduce to the ones we study here in the limit of a membrane in which spatial variations occur along 
one direction only.  

We find that there are two principal effects of curvature. The first is the opening of a frequency gap in the dispersion relation. 
This prevents bending $f$-modes with frequencies $\omega<M$, with $\omega$ and $M$ being the dimensionless 
frequency and curvature respectively. This is the simpler one-dimensional equivalent of the suppression of 
bending undulations on membranes at areas of positive Gauss curvature~\cite{vaziri2008localized,evans2013reflection}. For a finite 
rod with discrete frequency spectrum, the restriction of $\omega>M$ for bending eigenfrequencies causes eigenfrequencies to cross 
with increasing curvature. By slowly bending a ringing rod, one can, in effect,  ``hear" the effects of curvature by noting the 
modes split into an upper branch tending to the curve $\omega =M$, and a lower branch tending to zero. In this restricted sense, one can 
indeed hear changing curvature in a rod.  We also note that one observes an oscillation of eigenfrequencies with respect to $M$, 
as consequence of level splitting among harmonics.

The second principal effect of curvature is the ability for undulatory $f$-waves with frequency $\omega$, to tunnel through
regions of curvature $M>\omega$. Though the curved region cannot support such bending waves, by coupling to in-plane modes, 
these undulatory waves can convert to compression waves in order to tunnel through curvature. This tunneling effect may be significant for 
understanding the propagation of flexural (bending) phonons over large distances in rods or membranes with complex curvature in their
stress-free state.  Physical examples should include the propagation of phonons in bent carbon nanotubes or ribbons, as well as the propagation of 
membrane undulations along cell membranes~\cite{evans2017geometric}. 

One may inquire if multiple scattering of bending waves from randomly curved surfaces can lead to localization, and then consider how the ``tunneling" of bending waves may affect this result.  Both questions are interesting future directions. 

\begin{acknowledgments}
The authors thank Louis Foucard and Alexander Serov for fruitful discussions. 
This work was supported in part by NSF-DMR-1709785.
\end{acknowledgments}

\bibliography{referencesFU1}

\end{document}